\documentstyle{mn} 
\title[OH in LMC] 
{Observations of ground-state OH in the Large Magellanic Cloud} 
\author[K. J. Brooks \& J. B. Whiteoak]
{K. J.~Brooks$^{1,2}$ 
and J. B.~Whiteoak$^2$\\
$^1$Astrophysics \&
Optics, The University of NSW, Sydney 2052, NSW, Australia; kbrooks@newt.phys.unsw.edu.au\\
$^2$Australia Telescope National Facility, CSIRO, PO Box 76, Epping, NSW 2121, Australia}

\date{Accepted
1997 June 12. Received 1997 May 2}

\pubyear{1997}

\begin{document}

\label{firstpage}
\maketitle

\begin{abstract} We have carried out a series of observations of the
1665- and 1667-MHz transitions of the $^{2}$$\Pi$$_{3/2}$, J=3/2 OH
ground state towards six selected HII regions in the Large Magellanic
Cloud (IRAS 05011-6815 and MRC 0510-689, 0513-694B, 0539-691,
0540-696B, 0540-697A) using the Australia Telescope Compact Array. The
study has provided the first accurate positions for known 1665- and
1667-MHz OH masers as well as detecting several new masers. The
regions all contain H$_{2}$O or CH$_{3}$OH masers but OH masers were
detected in only four. The 1.6-GHz continuum emission was also imaged
to investigate its spatial relationship to the associated OH
maser. Although some masers are close to compact continuum
components, in other cases they are near the continuum
distribution boundaries and perhaps have been created as a result of
the HII region interacting with the surrounding interstellar medium.
   
\end{abstract}


\section{introduction}

The interstellar hydroxyl radical (OH) is ubiquitous in the molecular
clouds of our Galaxy, and its spectral-line transitions have been widely studied over the last 30
yr. Observations of the 1.6-GHz ground-state
$^{2}$$\Pi$$_{3/2}$, J=3/2 transitions have proven valuable in
investigations of the physical and kinematic properties of the
associated molecular
clouds. OH maser emission found in or near many giant molecular
clouds provides pointers for locations where star formation is
occurring or possibly about to occur.

Attempts at similar studies of the Magellanic Clouds have been limited
because little OH has been detected. In spite of several OH searches,
previous detections, all in the Large Magellanic Cloud (LMC),
consisted of maser emission near two HII regions and OH
absorption against the continuum emission of two other HII regions
(Whiteoak \& Gardner 1976; Caswell \& Haynes 1981; Haynes \& Caswell
1981; Gardner \& Whiteoak 1985; Caswell 1995). The OH deficiency may
be related to the lower metallicity and stronger UV-radiation fields
in the clouds compared with our Galaxy.

A limited number of other LMC masers have now been detected in
22-GHz transitions of H$_{2}$O (Whiteoak et
al. 1983; Whiteoak \& Gardner 1986) and 6-GHz transitions of
CH$_{3}$OH  (Sinclair et al. 1992; Ellingsen et al. 1994; Beasley et
al. 1996). Except for Caswell's (1995) observations,
the successful OH observations were obtained using the Parkes 64-m
radio telescope with a beamwidth of 12.5 arcmin. The angular
resolution plus uncertain positional results have hindered comparison
of the OH results with those for the other masers. To further our
knowledge of the OH detections, and to extend the search to the
locations of the other molecular masers, we have carried out a
series of observations of the 1665- and 1667-MHz transitions of the
OH ground state, using the Australia Telescope Compact Array (ATCA),
operated by the Australia Telescope National Facility, CSIRO.  The
observations also yielded the distribution of associated 1.6-GHz continuum
emission.

\section{observations}

The ATCA observations were carried
out in several periods between 1994 March and 1996 November.
Details of the
instrument are given by Frater \& Brooks (1992). The antenna spacings
ranged from 153 m to 6 km. The
observations and data processing were made using standard
procedures. An observing cycle was adopted in which 40-min target
observations were bracketed by 3-min observations of phase calibrator
PKS 0823-500; in some cases two different targets shared the period
between calibrator observations.  This calibrator was also used to
calibrate the spectral bandpass.  Flux-density was calibrated by observing PKS 1934-638 (adopted as the ATCA primary
calibrator), for which a flux density of 14.16 Jy was adopted.  The
correlator configuration consisted of 2048 channels extending over a
4-MHz band centered at 1665 MHz. This provided a channel spacing of
1.95 kHz (equivalent radial velocity of 0.35 km s$^{-1}$).  Any 1665- and
1667-MHz transitions of OH in the LMC would be expected to appear in this band.

 The correlated spectral outputs from pairs of ATCA antennas were
processed mostly at the Paul Wild Observatory, Narrabri using a
package based on the Astronomical Image Processing System (AIPS)
produced by the US National Radio Astronomy Observatory.  After
amplitude, phase and bandpass calibration, the linearly polarized
outputs were combined into total intensities.  Contributions of
continuum emission were removed from the spectral-line channels and
stored in a separate database.  The spectral-line database was split
into databases for each OH transition, and the frequency scales
converted into scales of heliocentric radial velocity.  Data in the
final databases were Fourier transformed, providing continuum and
spectral-line images which were `cleaned' to remove side-lobe fringes
caused by using only a limited number of baselines.  In this process,
`uniform' weighting was used for the imaging because it yielded the
highest angular resolution.  However, in some cases `natural'
weighting was used to image the continuum emission because this is
more effective in defining the faint extended emission.

 The selected targets included HII regions in which OH had been
previously detected by Whiteoak \& Gardner (1976), Caswell \& Haynes
(1981), Haynes \& Caswell (1981), and Gardner \& Whiteoak (1985): MRC
0510-689 (MC23, N105A), MRC 0539-691 (MC74, N157A), MRC 0540-696B (MC76, N160A), and MRC 0540-697A (MC77, N159).  It also included MRC
0513-694B (MC24, N113C), an HII region associated with the brightest
H$_{2}$O maser found in the LMC (Whiteoak \& Gardner 1986), and IRAS
05011-6815, a suspected ultra-compact HII region associated with a
bright CH$_{3}$OH maser (Beasley et al. 1996).

\section{results}

OH maser emission or thermal absorption has been
detected near all regions, except MRC 0539-691, associated with the
Doradus nebula.  In this case, the absorption detected with the Parkes
radio telescope is presumably extended, with a surface density too low
to be detected with the ATCA.  Table 1 summarizes the OH results for
the detected cases. The positional errors are given in parentheses. A
discussion of the individual results follows.

\subsection{IRAS 05011-6815}

A CH$_{3}$OH maser was detected towards this region by Beasley
et al. (1996). It showed three components located at RA(J2000) =
05$^{h}$01$^{m}$01$^{s}$.85, Dec(J2000) =
-68$^{\circ}$10$\arcmin$28$\arcsec$.3, with heliocentric velocities in
the range 265.6 - 267.9 km s$^{-1}$. The object had been selected from the
IRAS Point Source Catalogue (1985) on the basis of the criteria
suggested by Wood \& Churchwell (1989) for ultra-compact HII (UCHII)
regions.

Both 1665- and 1667-MHz OH maser emission was detected in images
created from the spectral database using a beam of extent 7.0 x 3.4
arcsec$^{2}$ (full width to half maximum).  Fig. 1a,b
shows the OH spectra of this emission.  The 1665-MHz spectrum consists
of a narrow feature at a heliocentric radial velocity of 267.6 km
s$^{-1}$ peaking at a flux density of 230 mJy beam$^{-1}$.  Its
measured position is RA(J2000) = 05$^{h}$01$^{m}$01$^{s}$.91
($\pm$0$^{s}$.03), Dec(J2000) = -68$^{\circ}$10$\arcmin$28$\arcsec$.5
($\pm$0$\arcsec$.2).  The 1667-MHz spectrum contains a narrow feature
at 266.2 km s$^{-1}$ with a peak flux density of 129 mJy beam$^{-1}$,
and possibly a fainter narrow feature at 268.7 km s$^{-1}$. The
position is similar to that of the 1665-MHz emission. The similarity of
OH and CH$_{3}$OH positions and velocities suggests a common
origin for the molecules. 

The IRAS position differs from the OH maser position by
1$^{s}$.06 in right ascension and -17$\arcsec$.8 in
declination. A 1.6-GHz continuum image made with a beam size of 12.3 x 5.8
arcsec$^{2}$ showed no evidence of any continuum emission towards or near
the maser, to an upper limit of 0.5 mJy beam$^{-1}$.  This is consistent with
the results of Beasley et al. (1966), who failed to detect any
continuum emission at 8.8 GHz.  

\subsection{MRC 0510-689 (MC23, N105A)}

H$_{2}$O and CH$_{3}$OH masers have already been detected in this HII
region, which has an H109$\alpha$ hydrogen recombination-line velocity
of 253 km s$^{-1}$ (McGee, Newton \& Brooks 1974).  The H$_{2}$O maser
was discovered by Scalise \& Braz (1981), and further observed by
Whiteoak et al. (1983) and Whiteoak \& Gardner (1986). Unpublished
observations by J. B. Whiteoak, T. Kuiper, P. Harbison and R-S Peng using
the 70-m antenna of NASA's Canberra Deep Space Communication Complex
(CDSCC) yield a preliminary position for the H$_{2}$O maser of RA(J2000) =
05$^{h}$09$^{m}$52$^{s}$.2, Dec(J2000) =
-68$^{\circ}$53$\arcmin$32$\arcsec$, and a velocity range of 253-268
km s$^{-1}$. The CDSCC positions should be
accurate to better than $\pm$3 arcsec. The CH$_{3}$OH
maser was discovered with the Parkes 64-m radio telescope by Sinclair
et al. (1992), and was subsequently re-observed at Parkes and imaged
with the ATCA by Ellingsen et al. (1994).  The emission, extending
over the velocity range 249-253 km s$^{-1}$, had a peak flux density
of 170 mJy. However, there was some uncertainty in its measured position
and we have reobserved the maser in 1997 March with the ATCA. The results
yielded a well-defined position RA(J2000) = 05$^{h}$09$^{m}$58$^{s}$.66 ($\pm$0$^{s}$.01), Dec(J2000) =
-68$^{\circ}$54$\arcmin$34$\arcsec$.1 ($\pm$0$\arcsec$.1). 1665-GHz OH maser emission in
the HII region was discovered at Parkes by Haynes \& Caswell (1981),
who determined a position with an rms uncertainty of 22 arcsec; it was
later re- observed by Gardner \& Whiteoak (1985).

Our ATCA maser images were produced with a beamwidth of 4.1 x 7.7 arcsec$^{2}$.
Both 1665- and 1667-GHz emission was detected; Figs 2a,b show the
spectra derived near the maser centre.  The 1665-MHz spectrum
essentially consists of a narrow feature with a peak flux density of
580 mJy beam$^{-1}$, centred at 253.4 km s$^{-1}$.  A weak feature of
flux density about 90 mJy beam$^{-1}$ may also be present at 255.8 km
s$^{-1}$.  The 1667-MHz spectrum contains a peak flux density of 248
Jy beam$^{-1}$, at a velocity of 254.2 km s$^{-1}$.  The 1665-MHz
narrow feature at 253.4 km s$^{-1}$ is centred at RA(J2000) =
05$^{h}$09$^{m}$51$^{s}$.94 ($\pm$0$^{s}$.02) , Dec(J2000)
-68$^{\circ}$53$\arcmin$28$\arcsec$.5 ($\pm$0$\arcsec$.3). Within
the errors, this is similar to the 1667-MHz maser emission position.

The distribution of continuum emission (Fig. 3) was imaged with a
restoring beam of size 10 x 6 arcsec$^{2}$ elongated at position
angle 101$^{\circ}$.  The maximum flux density of 19 mJy  beam$^{-1}$ occurs in
a compact region centred at RA(J2000) = 05$^{h}$09$^{m}$52$^{s}$.86
($\pm$0$^{s}$.05), Dec(J2000) = -68$^{\circ}$53$\arcmin$01$\arcsec$.0
($\pm$0$\arcsec$.1). The OH maser emission is not associated with any
continuum maxima, but is on the southern edge of the continuum
distribution, perhaps where the HII region is interacting with the
interstellar medium. The H$_{2}$O maser has a similar
location but the CH$_{3}$OH maser is offset by more than 1 arcmin and
is significantly south of the continuum distribution.

\subsection{MRC 0513-694B (MC24, N113)}

This HII region was selected because it contains by far the brightest
H$_{2}$O maser in the LMC; in 1984 the line profile contained two
narrow features with peak flux densities of about 22 Jy and
velocities of 247.3 and 254.5 km s$^{-1}$ (Whiteoak \& Gardner 1986).
The detected maser emission covers the velocity range 238-258 km
s$^{-1}$. The preliminary CDSCC position is RA(J2000) =
05$^{h}$13$^{m}$25$^{s}$.3, Dec(J2000) =
-69$^{\circ}$22$\arcmin$44$\arcsec$.

Maser images formed with a beamsize of 6.9 x 5.5 arcsec$^{2}$
yielded well-defined 1665-MHz maser emission.  Its spectrum (Fig. 4)
shows a narrow feature at a velocity of 248.3 km s$^{-1}$ with a peak flux
density of 257 mJy beam$^{-1}$.  The emission is centred at RA(J2000) =
05$^{h}$13$^{m}$25$^{s}$.18 ($\pm$0$^{s}$.05), Dec(J2000) =
-69$^{\circ}$22$\arcmin$46$\arcsec$.0 ($\pm$0$\arcsec$.1)

Fig. 5 shows the distribution of 1.6-GHz continuum emission imaged
with a 9.6 x 6.3 arcsec$^{2}$ beam elongated at a position angle of 6.9$^{\circ}$.  Superimposed on faint extended emission are
three regions centred at positions RA(J2000) =
05$^{h}$13$^{m}$17$^{s}$.74 ($\pm$0$^{s}$.03), Dec(J2000) =
-69$^{\circ}$22$\arcmin$24$\arcsec$.0 ($\pm$0$\arcsec$.1), RA(J2000) =
05$^{h}$13$^{m}$21$^{s}$.65 ($\pm$0$^{s}$.04), Dec(J2000) =
-69$^{\circ}$22$\arcmin$39$\arcsec$.9 ($\pm$0$\arcsec$.1), RA(J2000) =
05$^{h}$13$^{m}$25$^{s}$.09 ($\pm$0$^{s}$.09), Dec(J2000) =
-69$^{\circ}$22$\arcmin$45$\arcsec$.4 ($\pm$0$\arcsec$.2) The
respective peak flux densities are 51, 31 and 10 mJy beam$^{-1}$.  The
regions are quite compact; their sizes, corrected for the imaging beam
shape, are in the range 3-7 arcsec.  An image obtained using natural
weighting shows that the faint emission extends in right ascension
from 05$^{h}$13$^{m}$00$^{s}$ to 05$^{h}$13$^{m}$50$^{s}$, and in
declination from -69$^{\circ}$24$\arcmin$ to -69$^{\circ}$16$\arcmin$.

In summary, the OH and
H$_{2}$O masers probably originate in a common cloud region associated
with the most eastern small-diameter continuum component.

\subsection{MRC 0540-696B (MC76, N160A)}

An H$_{2}$O maser was detected in this HII region (Whiteoak et
al. 1983; Whiteoak \& Gardner 1986) which has an H109$\alpha$
recombination-line velocity of 254 km s$^{-1}$ (McGee et al.  1974).
The preliminary CDSCC position of the maser is RA(J2000) =
05$^{h}$39$^{m}$43$^{s}$.7, Dec(J2000) =
-69$^{\circ}$38$\arcmin$31$\arcsec$. Caswell \& Haynes (1981)
discovered a 1665-MHz OH maser which they believed could be associated
with the H$_{2}$O maser that Scalise \& Braz (1981) had reported in
N159.  However, later OH observations by Gardner \& Whiteoak (1985)
supported a location of the OH maser in MRC 0539-696B.  Imaging the
ATCA spectral data with a restoring beam of width 7.6 x 2.8
arcsec$^{2}$ confirmed the association of the maser with this HII region.

Fig. 6 shows the spectrum of the maser emission.  A narrow feature
with a peak flux density of 221 mJy beam$^{-1}$ is present at a
velocity of 248.6 km s$^{-1}$. A fainter feature of amplitude about 80
mJy beam$^{-1}$ is centred at about 244.5 km s$^{-1}$. The position
determined for the stronger feature is RA(J2000) =
05$^{h}$39$^{m}$39$^{s}$.0 ($\pm$0$^{s}$.5), Dec(J2000) =
-69$^{\circ}$39$\arcmin$11$\arcsec$.1 ($\pm$0$\arcsec$.1) Within the
errors of measurement, the weaker feature is located at the same
position. There was no detection of 1667-MHz emission using the same
beam size to an upper limit of 25 mJy.

Caswell (1995) recently detected maser emission in ATCA 
observations of the 6035-MHz, excited-state, OH transition centred at a
position offset from the 1665-GHz OH position by -0$^{s}$.08 in right
ascension and 0$\arcsec$.1 in declination.  The position coincidence
suggests a common origin for the two masers.

Fig. 7 shows the distribution of the 1.6-GHz continuum emission,
imaged with a circular restoring beam of  6-arcmin diameter.
Superimposed on a faint extended emission is a small-diameter
component centred at RA(J2000) = 05$^{h}$39$^{m}$46$^{s}$.15
($\pm$0$^{s}$.5), Dec(J2000) = -69$^{\circ}$38$\arcmin$39$\arcsec$.4
($\pm$0$\arcsec$), with peak flux density of 105 mJy beam$^{-1}$.  The OH
maser is not associated with this feature, but with a fainter continuum 
component about 50 arcsec to the south-west.  The H$_{2}$O maser position is
about 15 arcsec north-west of the main continuum component and,
despite the significant possible errors in the position estimates,
cannot be associated with the OH maser.

\subsection{MRC 0540-697A (MC77, N159)}

This HII region is one of the brightest in the LMC, and yielded the
first detection of interstellar OH (in absorption) in this galaxy
(Whiteoak \& Gardner 1976).  ATCA observations of 5-GHz spectral-line
and continuum distributions (Hunt \& Whiteoak 1994) revealed a second
compact radio component that is totally obscured at optical
wavelengths by a dense interstellar cloud. Subsequent mm-wavelength
observations (e.g. Chin et al. 1997) have established this cloud as
one of the best target candidates for molecular-line studies of the
LMC. Scalise \& Braz (1981) claimed to detect an H$_{2}$O
maser in this HII region, but subsequent attempts to confirm the
detection were unsuccessful (see e.g. Whiteoak \& Gardner 1986).
Finally, a maser was detected beyond the south-western boundary of the
HII region (unpublished observations made with the Parkes 64-m
radio telescope). The preliminary CDSCC position is RA(J2000) = 05$^{h}$39$^{m}$31$^{s}$.2, Dec(J2000) = -69$^{\circ}$47$\arcmin$28$\arcsec$.
 
Fig. 8 shows the distribution of 1.6-GHz continuum emission obtained
with the ATCA and imaged with a circular restoring beam of 6-arcmin diameter.  The maximum flux density of 91 mJy beam$^{-1}$ is associated with
the optically obscured compact continuum component, centred at
RA(J2000) = 05$^{h}$39$^{m}$37$^{s}$.60 ($\pm$0$^{s}$.02), Dec(J2000)
= -69$^{\circ}$45$\arcmin$25$\arcsec$.9 ($\pm$0$\arcsec$.1).

Imaging of the OH data revealed no maser emission, but a 1667-MHz
absorption feature (Fig. 9) was found to be centred at RA(J2000) =
05$^{h}$39$^{m}$37$^{s}$.60 ($\pm$0$^{s}$.02), Dec(J2000) =
-69$^{\circ}$45$\arcmin$25$\arcsec$.9 ($\pm$0$\arcsec$.1),
i.e. coincident with the compact continuum component.  The feature
has a central velocity of 254.9 km s$^{-1}$ and a maximum line-to-continuum ratio of -0.30.  Efforts to detect OH absorption towards
the fainter compact region further to the east yielded an upper
limit of -0.15 in line-to-continuum ratio.  Clearly the results
support the existence of a dense molecular cloud over the western side
of the HII region.  This is the only ATCA detection of OH absorption in
the HII regions observed in this survey.

\section{conclusions}

This study has provided the first accurate positions for 1665- and
1667-MHz OH masers in the LMC, and reveals the relationship between
the masers and associated HII regions. In three of the four cases (out
of the six selected HII regions) in which OH masers were found, their
positions were very close to those of other masers; in the fourth, the
positions differed considerably.  Although some masers were close to
compact continuum components, others were near the continuum
distribution boundaries and may have been created as a result of the
HII region interacting with the interstellar medium.

The flux densities of the 1665-MHz masers ranged
between 221 and 580 mJy; for an assumed LMC distance of 55 kpc, the
equivalent range in `luminosity' is 669 to 1754 Jy kpc$^{2}$. Our
detection limit was equivalent to a luminosity of 150 Jy kpc$^{2}$. In
the OH maser statistics of our Galaxy (Caswell \& Haynes 1987), 50
per cent of the masers with intensities above this threshold have
intensities equal to or greater than the detected LMC masers. It
should be possible to detect additional masers in future surveys of
the LMC. An unbiased
survey of HII regions would be required to provide firm statistical trends
for comparison with a survey made in our Galaxy.

\section{Acknowledgments}

The first detection of the OH maser near IRAS 05011-6815 was made by
Dr J. L. Caswell who generously passed his information on to the
authors.

\newpage

\section{Captions to figures:}

Fig. 1: Spectra of (a) 1665-MHz and (b) 1667-MHz OH maser emission
near IRAS 05011- 6815.  

\noindent Fig. 2: Spectra of (a) 1665-MHz and (b) 1667-MHz OH maser emission in
MRC 0510-689.  

\noindent Fig. 3: Distribution of 1.6-GHz continuum emission of MRC 0510-689.  Contour levels are 0.02,
0.05, 0.1, 0.15, 0.2, 0.25, 0.3, 0.4, 0.5, 0.6, 0.7 units of 27.8
mJy beam$^{-1}$. The cross and triangle mark OH and H$_{2}$O maser
positions, respectively. The position of the nearest CH$_{3}$OH maser
is 45 arcsec south of the southern limit in
the figure.

\noindent Fig. 4: Spectrum of 1665-MHz OH maser emission in MRC 0513-694B.

\noindent Fig. 5: Distribution of 1.6-GHz continuum emission of MRC 0513-694B.  Contour levels are 0.05,
0.1, 0.15, 0.2, 0.3, 0.4, 0.5, 0.6, 0.7, 0.8, 0.9, 1.0 units of the
peak flux density of 51.5 mJy beam$^{-1}$.  The cross and triangle mark OH and
H$_2$O maser positions, respectively.

\noindent Fig. 6: Spectrum of 1665-MHz OH maser emission in MRC
0540-696B.

\noindent Fig. 7: Distribution of 1.6-GHz continuum emission of MRC 0540-696B.  Contour levels are 0.02, 0.05, 0.1,
0.2, 0.3, 0.4, 0.5, 0.6, 0.7, 0.8, 0.9, 1.0 units of the peak flux
density of 105 mJy beam$^{-1}$.  The cross, triangle and star mark OH,
H$_{2}$O and 6035-MHz excited state OH maser positions, respectively.

\noindent Fig. 8: Distribution of 1.6-GHz continuum emission of MRC
0540-697A.  Contour levels are 2, 5, 10, 20, 30, 40, 50, 60, 70, 80
and 90 mJy beam$^{-1}$.

\noindent Fig. 9: Spectrum of 1667-MHz OH absorption against the brightest
compact component of MRC 0540-697A.

\newpage

\begin{table*}
\begin{minipage}{210mm}
\caption{Summary of OH Spectra Parameters} 
\begin{tabular}{@{}lllllll}

Region		& Recomb& OH line& RA(2000)		& DEC(2000)
	& Peak  & Peak \\
		& Velocity&      &			&
	& Velocity & Intensity\\ 
		& km s$^{-1}$&MHz& ~h ~~m ~~s		& ~~\degr  ~~$'$ ~~\arcsec         
		& km s$^{-1}$ & mJy beam$^{-1}$\\

\hline 
\\
IRAS 05011	& 	& 1665	& 05 01 01.91 (0.03)	& -68 10 28.5
(0.2)	& 267.6	& 230\\ 
		&	& 1667	& 05 01 01.91 (0.08)	& -68 10 28.1 (0.4)
	& 266.2 & 129\\
\\ 
MRC 0510-689	&  253	& 1665	& 05 09 51.94 (0.02)	& -68 53 28.5
(0.2)	& 253.4 & 580\\ 
(MC23, N105a)	&       & 1665  & 05 09 52.18 (0.12)	& -68 53 28.7 (0.2)
	& 255.8	& 90\\
		&	& 1667	& 05 09 52.00 (0.03)	& -68 52 28.6 (0.1)
	& 254.2 & 248\\ 
\\
MRC 0513-694B	& 	& 1665	& 05 13 25.18 (0.05)	& -69 22 46.0
(0.1)	& 248.3 & 257\\
(MC24, N113)\\
\\
MRC 0540-696B	& 254	& 1665 & 05 39 39.00 (0.5)	& -69 39 11.1
(0.1)	& 248.6 & 221\\
(MC76, N160A)	&	& 1665 & 05 39 39.15 (1)	& -69 39 10.9 (2)
	& 244.5	& 80\\
\\
MRC 0540-697A	& 254	& 1667 & 05 39 37.60 (0.02)	& -69 45 25.9
(0.1)	& 254.9 & -0.3 \footnote{line / continuum }\\
(MC77, N159)\\
 \\
\hline 
\end{tabular} 
\end{minipage}
\end{table*}

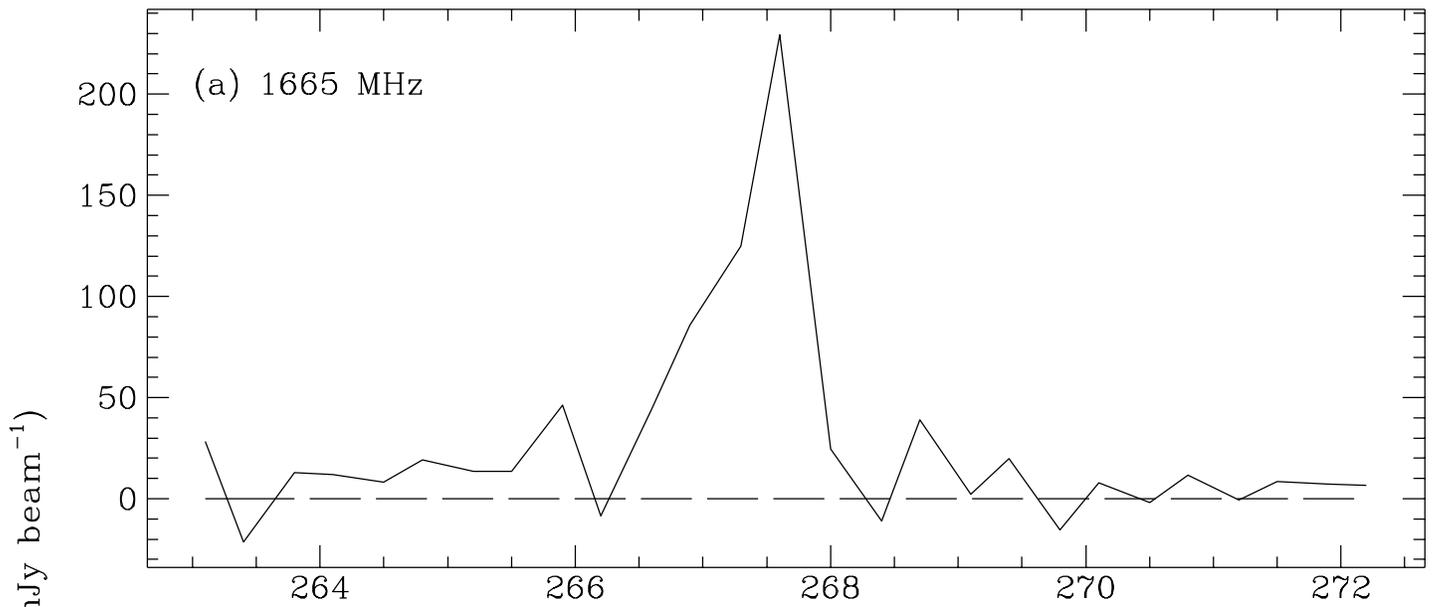
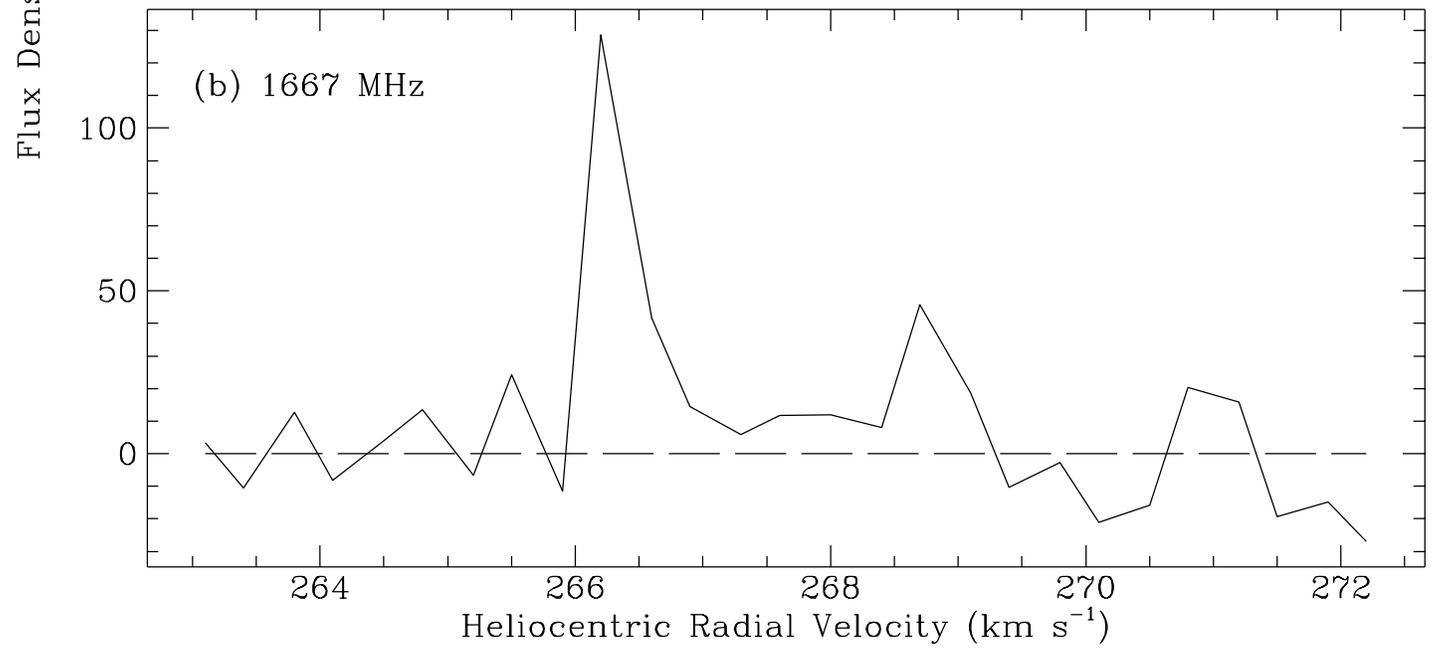
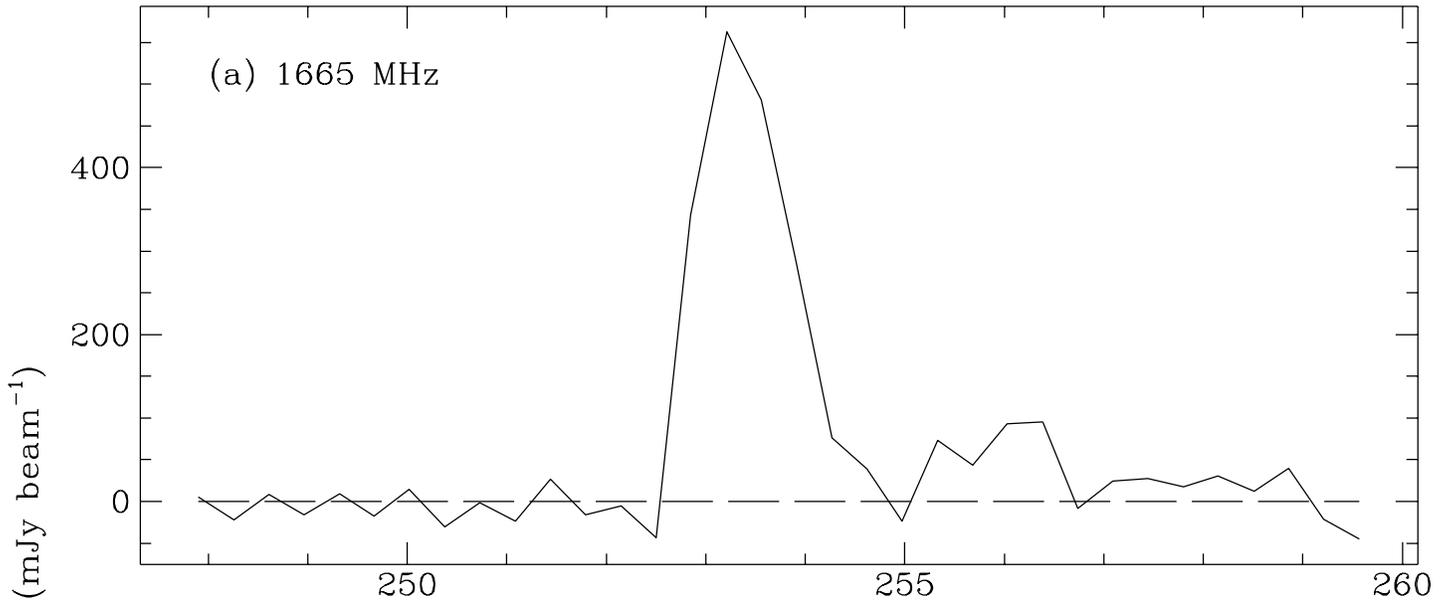
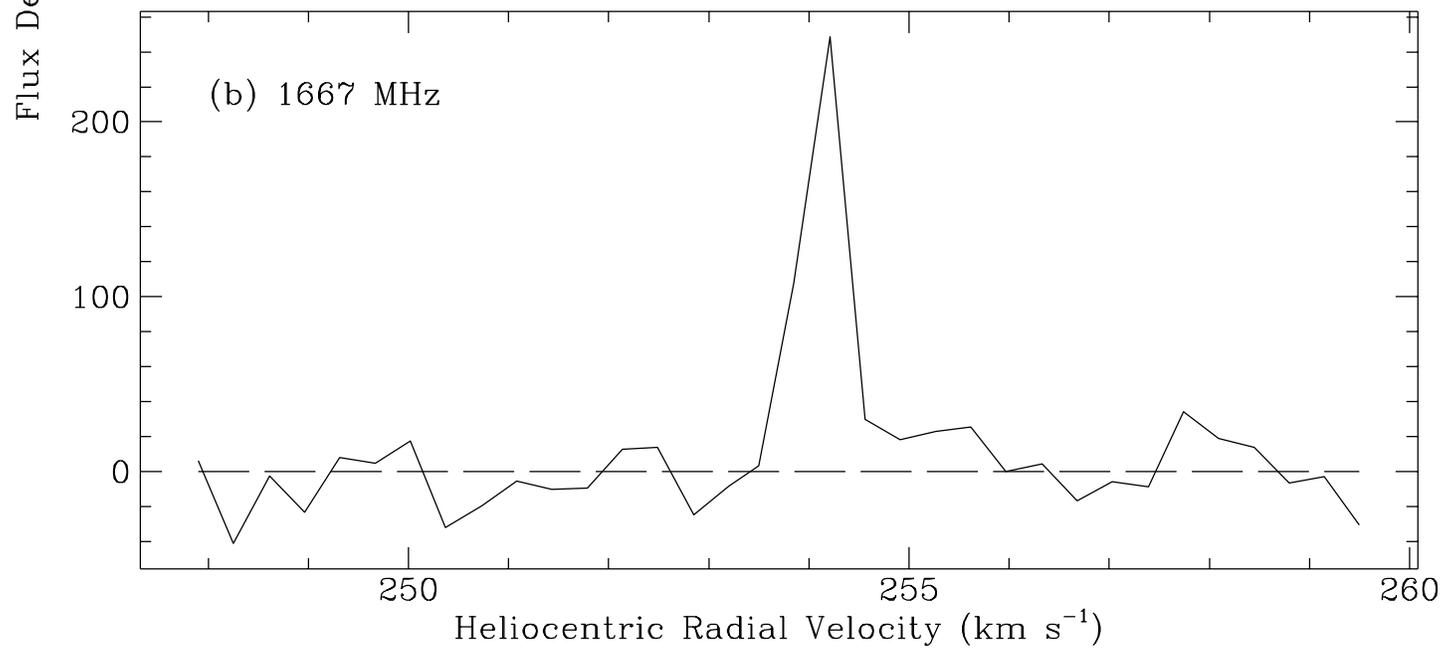
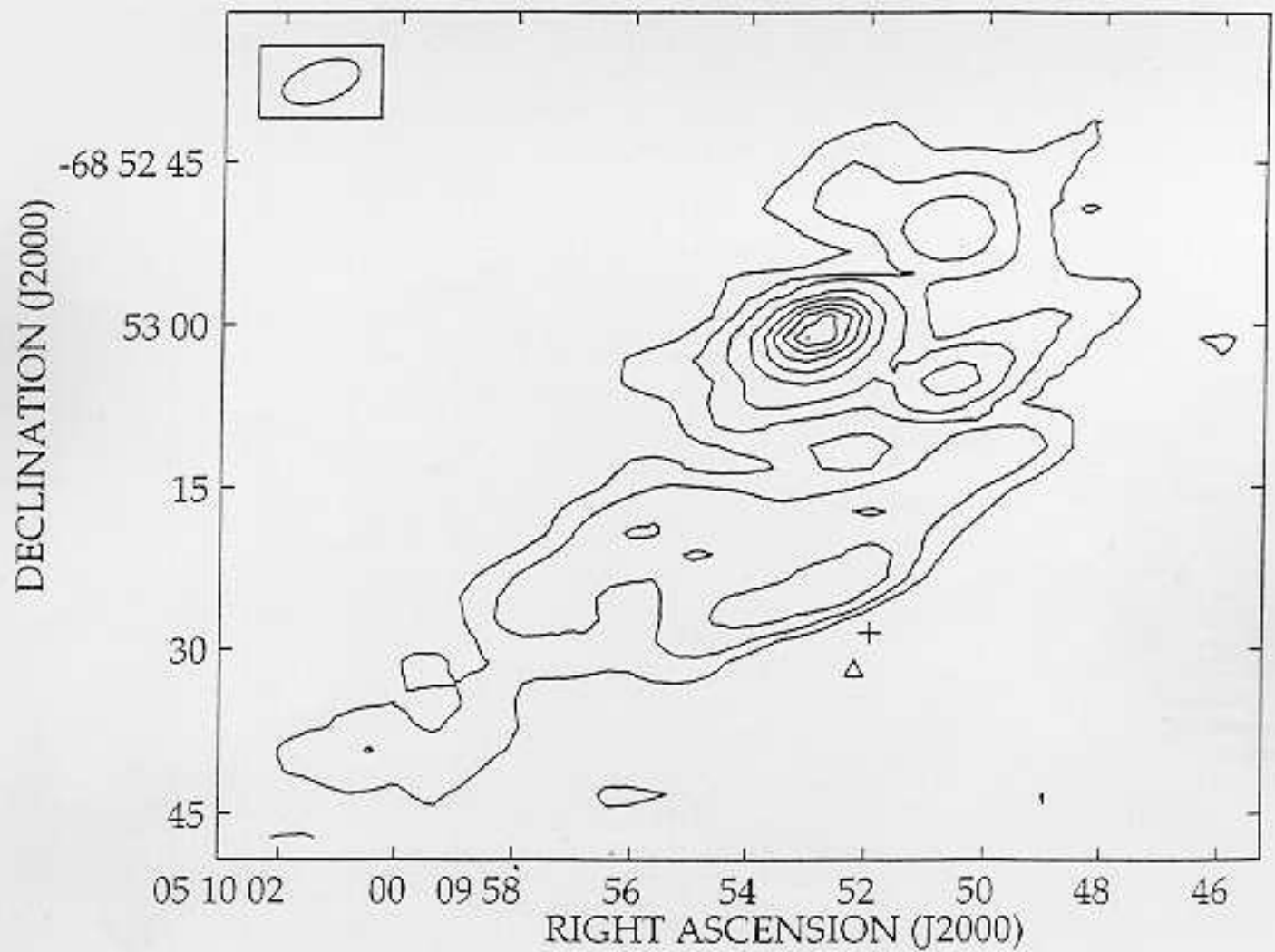
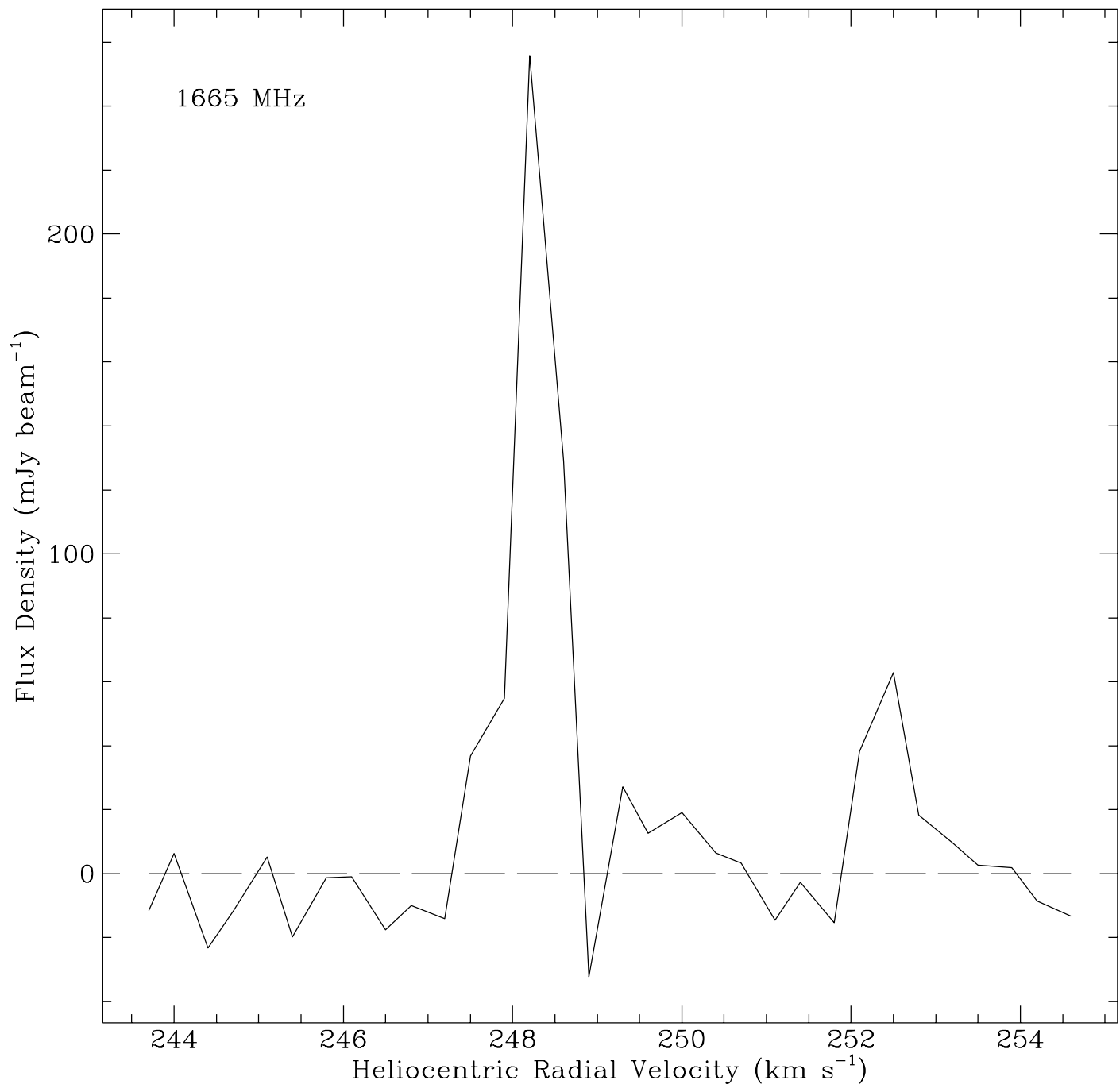
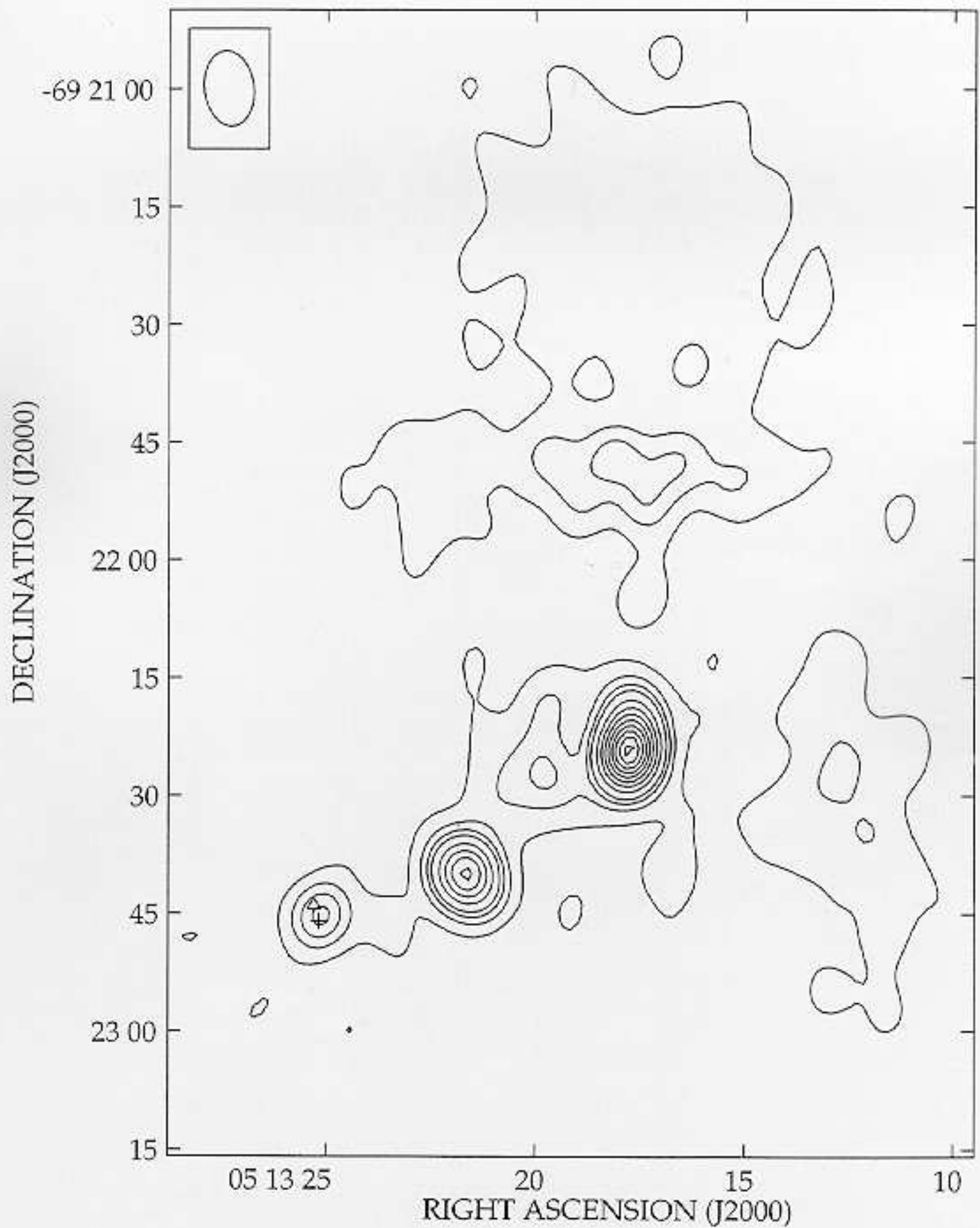
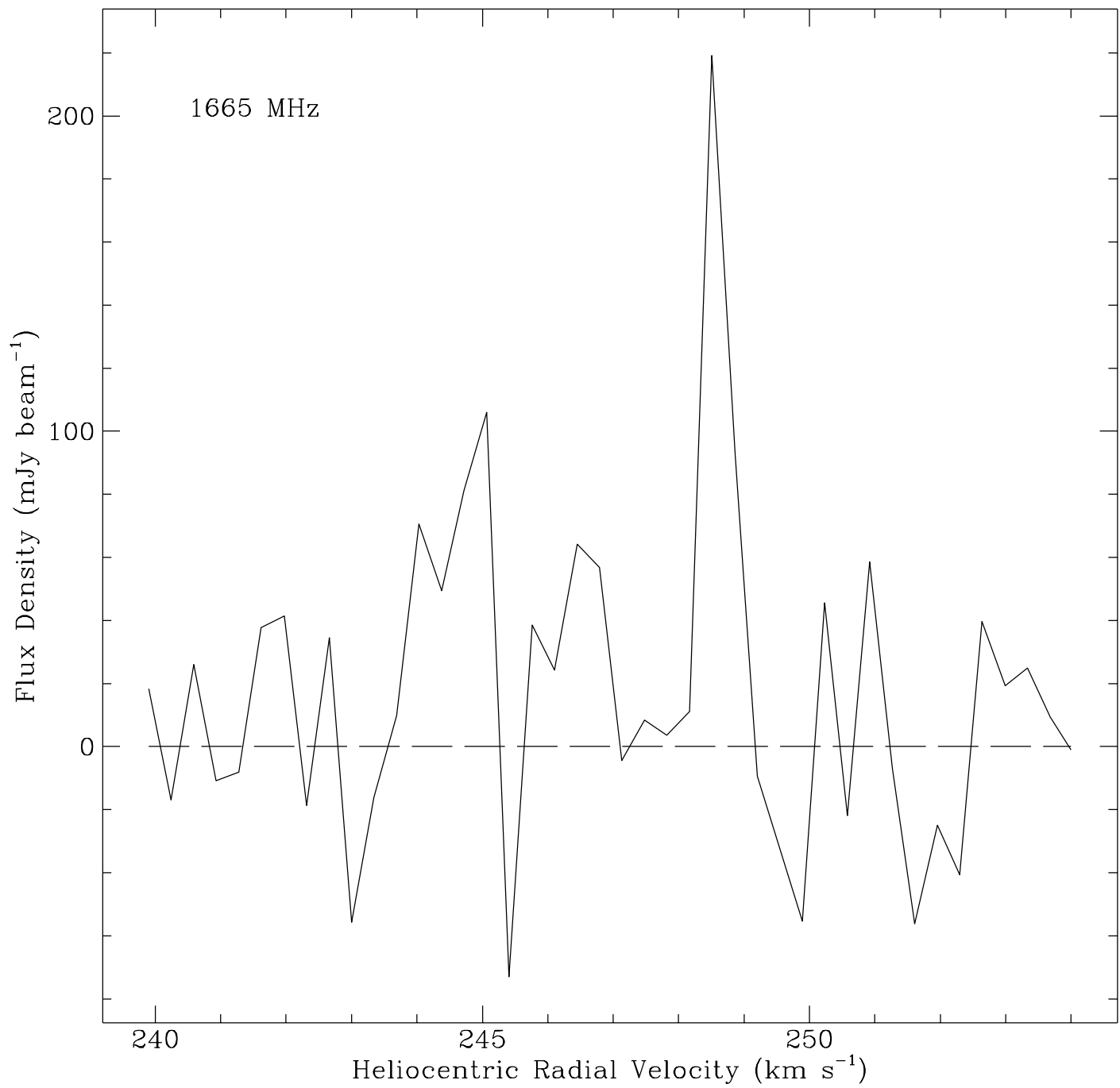
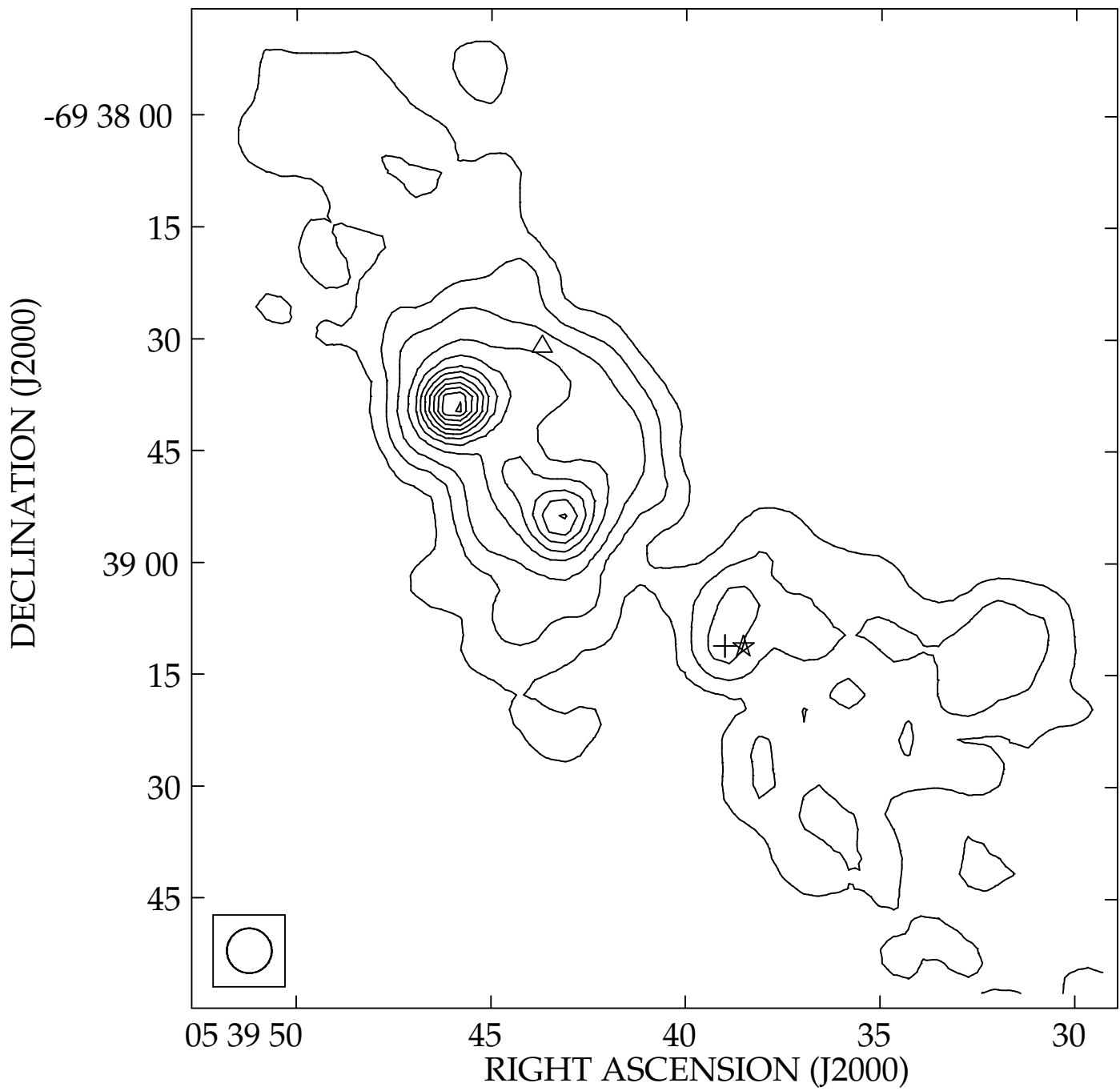
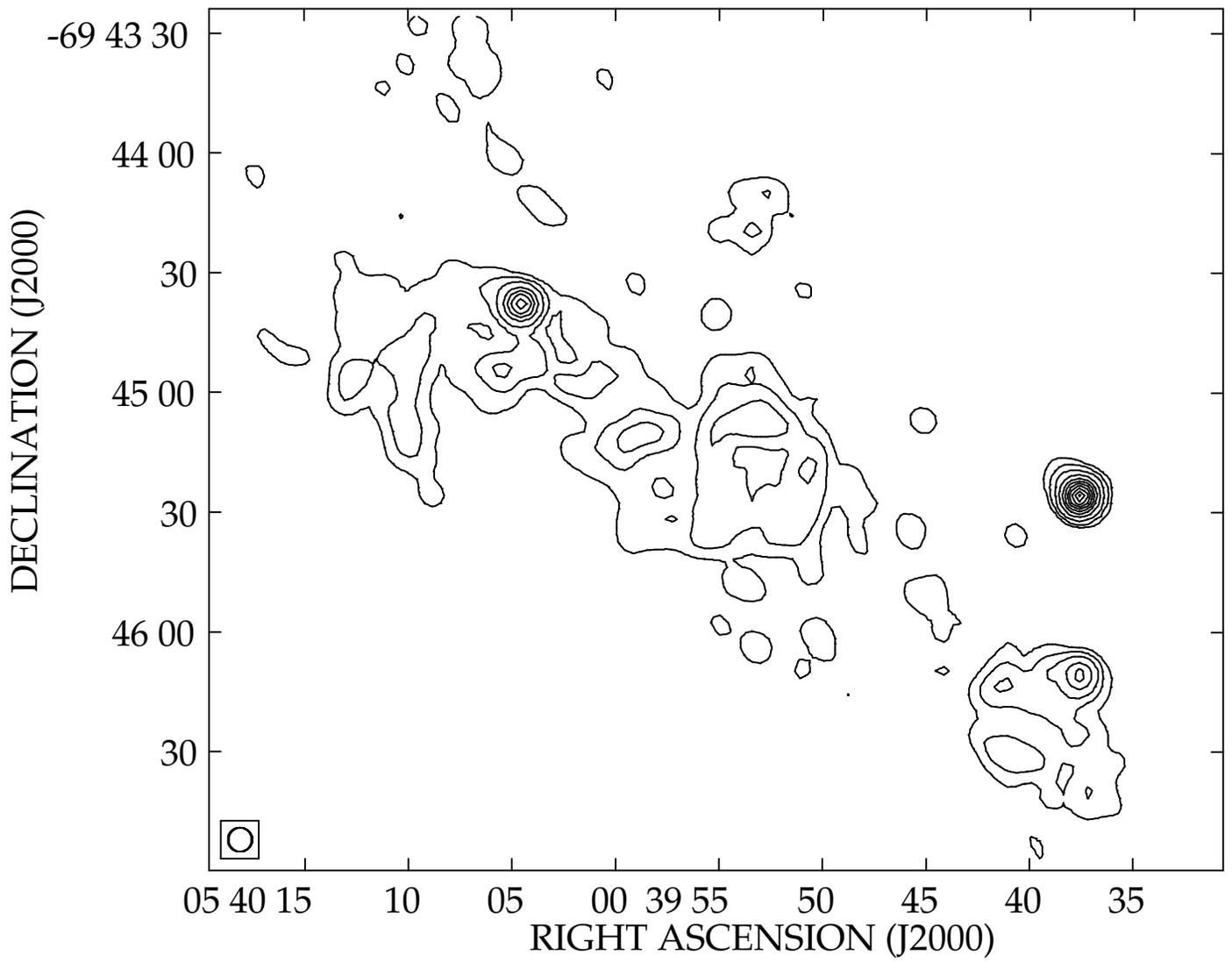
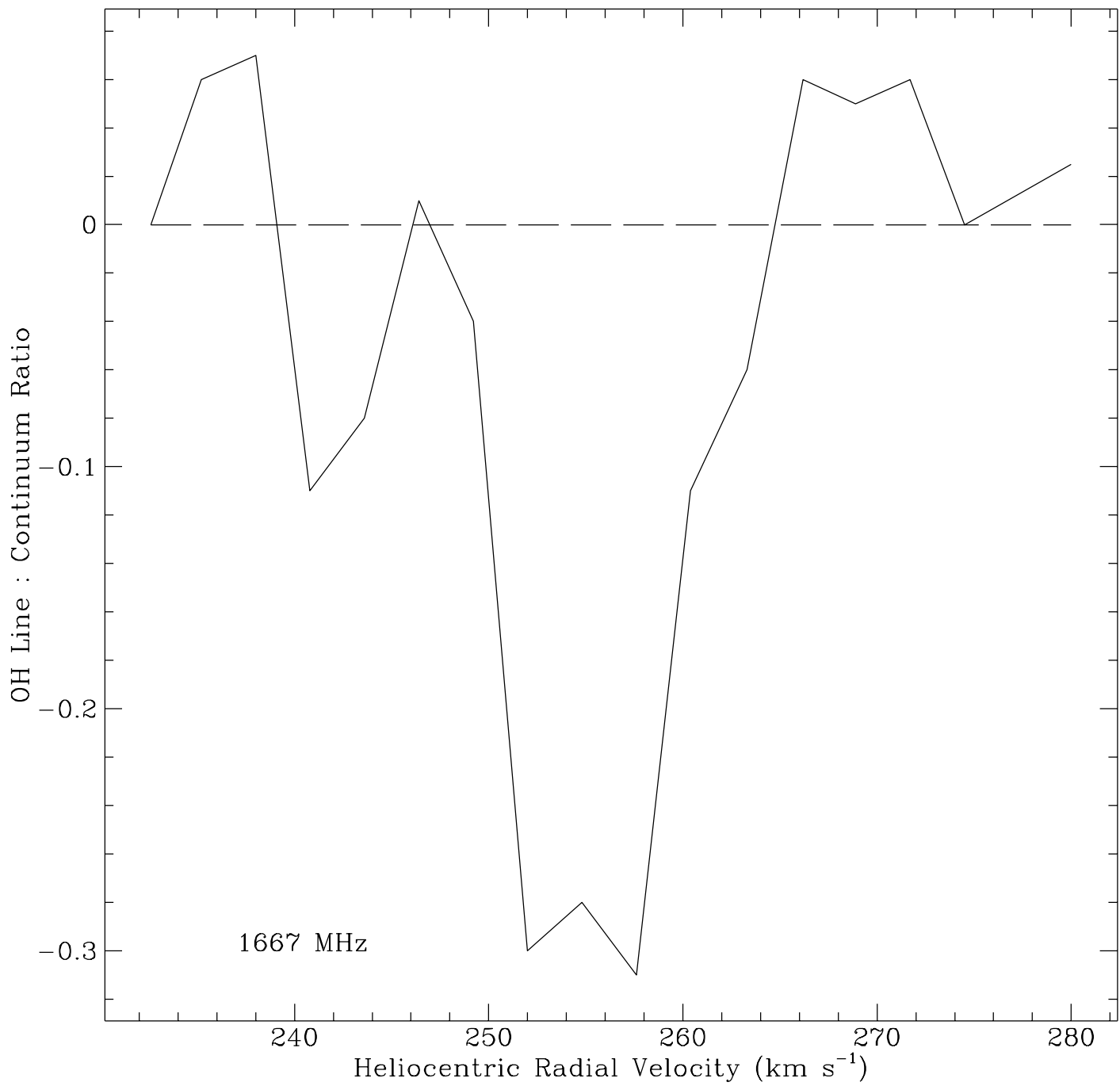
\end{document}